\documentclass[a4paper,11pt]{article}
\usepackage{jheppub} % for details on the use of the package, please
\usepackage{amsmath}
\usepackage{slashed}
\usepackage{xcolor}
\usepackage{graphicx}
\usepackage[T1]{fontenc} % if needed
\usepackage{booktabs}
\usepackage{makecell,comment}
\usepackage{siunitx}
\usepackage{titlesec}

\titleformat*{\subsection}{\normalfont\large\bfseries\boldmath}

\usepackage{cancel}

\usepackage{mathtools}

\input{defs}

\allowdisplaybreaks

\makeatletter
\divide\@tempdima\baselineskip
\@tempcnta=\@tempdima
\makeatother

\title{Bosonic Contribution to \\ Parity-Violating M\o ller Scattering at NNLO}

\author[a]{Lisong Chen,}
\author[a]{Jens Erler,}
\author[b]{Ayres Freitas,}
\author[b]{Juhun Kwak}

\affiliation[a]{{\rm PRISMA}$^{++}$ Cluster of Excellence and Institute for Nuclear Physics, \\
Johannes Gutenberg-University, 55099 Mainz, Germany}
\affiliation[b]{Pittsburgh Particle-physics Astro-physics \text{\&} 
Cosmology Center (PITT-PACC), \\
Department of Physics \text{\&}  Astronomy, University of Pittsburgh, Pittsburgh, PA 15260, USA \\ }

\emailAdd{lichen@uni-mainz.de}
\emailAdd{erler@uni-mainz.de}
\emailAdd{afreitas@pitt.edu}
\emailAdd{juhun.kwak@pitt.edu}

\abstract{We calculate the dominant bosonic two-loop virtual electroweak corrections to 
parity-violating Møller scattering at low energies.  
The electron effective weak charge is extracted from the polarization asymmetry, 
retaining all contributions that are enhanced by  $(1-4\sin^2\theta_W)^{-1}$ or 
$\ln^2(M_{W,Z}^2/Q^2)$ relative to the tree-level contribution. 
Analytical results are obtained using the expansion by regions method in the low-energy 
limit $Q^2\ll M_{W,Z}^2$.  
Within this approximation, the dominant two-loop bosonic corrections shift the 
electron effective weak charge (or equivalently the polarization asymmetry) by $0.3\%$.  
We examine the origin of the leading contributions to this result and assess the 
remaining theoretical uncertainty from purely bosonic contributions for the MOLLER 
experiment to about 0.1\%.}

\begin{document} 
\maketitle

\flushbottom

\section{Introduction}
\label{sec:introduction}
The Standard Model (SM) of particle physics is successful at describing Nature at 
the smallest scale currently accessible at accelerators and colliders. 
All of the elementary particles predicted by the SM have been discovered.
Some predictions such as the electron and muon anomalous magnetic moments have been 
tested to extremely high precision. 
However, many open questions and the ever growing evidence for dark matter (which cannot 
be described within the SM) call for physics Beyond the Standard Model (BSM). 
One can test BSM theories either directly by producing the corresponding BSM particles
in high-energy or high-intensity experiments, or through precision measurements of  
electroweak processes.

A powerful probe in the context of the latter is parity-violating electron scattering 
(PVES)~\cite{Erler:2014fqa} where otherwise identical beams of longitudinally polarized 
electrons of left-handed (L) or right-handed (R) helicities scatter from a fixed target.
%The cross-section asymmetry $A_{LR}$ is defined as
%\begin{equation}
%    A_{LR} \equiv \frac{d\sigma_L - d\sigma_R}
%    {d\sigma_L + d\sigma_R}.
%    \label{eqn:LRAsy}
%\end{equation}
Historically, the Glashow-Weinberg-Salam 
theory~\cite{Glashow:1961tr,Weinberg:1967tq,Salam:1968rm} was recognized as the best 
model over its alternatives through the measurement of the cross-section asymmetry $A_\text{LR}$,
defined as
\begin{equation}
    A_\text{LR} \equiv \frac{d\sigma_\text{L} - d\sigma_\text{R}}
    {d\sigma_\text{L} + d\sigma_\text{R}},
    \label{eqn:LRAsy}
\end{equation}
in deep-inelastic electron-deuteron scattering~\cite{Prescott:1979dh}. 
This experiment measured the weak mixing angle $\theta_W$ with a value in agreement with
those found in neutrino scattering experiments measuring the ratio of neutral-to-charged 
current events.
Measurements of atomic parity violation (APV) also fell in line.

The polarization asymmetry \eqref{eqn:LRAsy} in M\o{}ller scattering provides one of the 
most precise measurement of the weak mixing angle at energies below 
the $Z$ pole~\cite{Kumar:2013yoa}.
It was first measured by the E158 Collaboration at SLAC in 
the mid-2000's~\cite{SLACE158:2005uay}, confirming with $6\sigma$ significance 
the predicted renormalization group running~\cite{Czarnecki:1998xc,Erler:2004in} of 
$\sin^2\theta_W$ between the $Z$-boson mass, $M_Z$, and the average momentum transfer 
squared of the experiment, $Q^2 \approx 0.026$~GeV$^2$ (the beam time averaged 
center-of-mass energy was $E_\text{CM} = 219$~MeV).
The precision of the M\o{}ller asymmetry will see another notable improvement soon, 
thanks to the Measurement Of Lepton-Lepton Electroweak Reaction (MOLLER) 
experiment~\cite{MOLLER:2014iki} currently under construction at Jefferson Lab. 
MOLLER aims to determine $A_\text{LR}$ in M\o{}ller scattering at $Q^2\approx 0.0056$~GeV$^2$ 
($E_\text{CM} = 106$~MeV) with $2.4\%$ uncertainty, comparable to determinations near
the $Z$ resonance from LEP, the SLC, the Tevatron and the LHC.
Within the SM, the MOLLER experiment is a precision test of the scale dependence of 
$\sin^2\theta_W$~\cite{Czarnecki:1998xc,Erler:2004in,Erler:2017knj,Erler:2024lds}. 
Furthermore, MOLLER will provide insights into BSM physics to which high-energy 
colliders are blind, such as models with a dark (very small couplings to SM fermions) and 
light (mass well below a GeV) $Z'$ boson~\cite{Davoudiasl:2012ag}. 
MOLLER is also particularly sensitive to the doubly-charged scalar bosons appearing in 
left-right symmetric models~\cite{Pati:1974yy}.

M\o{}ller scattering benefits from a clean leptonic environment. 
Moreover, the leading-order (LO) asymmetry resulting from the interference of photon 
and $Z$ boson exchange amplitudes is proportional to the numerically small factor  
$1-4\sin^2\theta_W < 0.1$: 
\begin{equation}
    A_\text{LR} = \frac{G_F s}{\sqrt{2}\pi\alpha}\,
    \frac{y(1-y)}{1+y^4+(1-y)^4}(\underbrace{1-4\sin^2\theta_W+\delta Q_W^e}_{Q_W^e}),
    \label{ALR}
\end{equation}
where $G_F$ is the Fermi constant, $\alpha$ the fine structure constant, and 
$y=-t/s=Q^2/s$ is the fractional energy transfer in terms of the usual Mandelstam 
variables.
Thus, $A_\text{LR}$ features an enhanced sensitivity to $\sin^2\theta_W$,
along with new (BSM) physics that is not accompanied by the same suppression factor.

We will refer to the quantity $Q_W^e$ as the {\em effective\/} weak charge of the 
electron since a residual dependence on kinematic details remains after the inclusion of 
higher-order effects (by contrast, a true charge would have be static and 
process-independent).
The term $\delta Q_W^e$ in Eq.~\eqref{ALR} collects the electroweak radiative
corrections~\cite{Czarnecki:1995fw} from both the numerator and the
denominator in Eq.~\eqref{eqn:LRAsy}.
Some terms in $\delta Q_W^e$ do not carry the $1-4\sin^2\theta_W$ factor, so that their 
relative impact is enhanced. 
Examples of such contributions include the fermion loops that enter
the aforementioned renormalization group running of $\sin^2\theta_W$,
and vertices and box diagrams with $W$-boson--fermion couplings. 
In total, such one-loop electroweak radiative corrections were found to be as large as 
$-40\%$~\cite{Czarnecki:1995fw}, but this conclusion depends crucially on which 
definition 
of the weak mixing angle is inserted into the tree-level expression Eq.~\eqref{ALR}.
Indeed, this calculation assumed the \MSbar\ definition $\sin^2\hat\theta_W(\mu)$ 
with the 't Hooft scale chosen as $\mu = M_Z$.
Alternatively employing the on-shell definition
\begin{equation}
    s^2_W\equiv \sin^2\theta_W^{\rm (on-shell)} \equiv 1 - \frac{M_W^2}{M_Z^2} = 0.22348 \pm 0.00010,
\label{sin2tonshell}
\end{equation}
one would find that the radiative corrections reduce the tree level result by 
an {\em additional\/} $30\%$.
These kinds of corrections would cast doubt on the convergence of the perturbative 
expansion.
However, the reason for their presence can be traced to inappropriate scale choices.
Both $\sin^2\hat\theta_W(M_Z)$ (explicitly) and $\sin^2\theta_W^{\rm (on-shell)}$
(implicitly) are quantities defined at the electroweak mass scale, while the relevant 
scale in Eq.~\eqref{ALR} is $Q^2 \ll M_Z^2$. 
This mismatch generates large spurious logarithms and renders the perturbative results 
unreliable. 
However, using the renormalization group equation for 
$\hat\theta_W$~\cite{Erler:2004in,Erler:2017knj} one may choose the \MSbar\ scale 
$\mu = \sqrt{Q^2}$, thereby re-summing the spurious logarithms to all orders.

The remaining next-to-leading order (NLO) radiative corrections may {\em a priori\/} also 
be sizable.
Individual contributions are enhanced by factors of $g^2/[4\pi^2(1-\sin^2\theta_W)]$ or $\ln(M_Z^2/Q^2)$
and indeed are quite large.
However, there are fortuitous cancellations keeping the NLO radiative corrections at 
the level of 2\%.

The fermionic\footnote{From this point on, ``fermionic'' refers to contributions 
from diagrams with at least one closed fermion loop, whereas diagrams without closed fermion loops 
are dubbed ``bosonic''.} next-to-next-to-leading order (NNLO) corrections are of similar 
size~\cite{Du:2019evk}. 
Even though the bosonic NNLO corrections are expected to be smaller compared to the 
fermionic ones (due to the absence of the enhancement in the latter where all the SM 
fermions contribute with the same sign), it is important to perform this calculation to
obtain a reliable interpretation of the upcoming $2.4\%$ measurement of $A_\text{LR}$ 
in terms of SM or BSM physics.

Here we report on an approximate computation of the NNLO contributions without closed 
fermion loops. 
Since we expect them to be smaller than those from diagrams with closed fermion loops, we 
focus on the subset of {\em enhanced\/} bosonic NNLO corrections, namely 
(\emph{i}) terms without the suppression factor $1-4\sin^2\theta_W$, and 
(\emph{ii}) terms that are proportional to the square of a large logarithm,  
$\ln^2(M_Z^2/Q^2)\sim 200$.

\section{Method}
\label{sec:mtd}
We calculate the left--right asymmetry by expanding Eq.~\eqref{eqn:LRAsy}
consistently through two-loop order.
The parity-violating numerator originates from weak interactions: at Born level,
its leading contribution is generated by $\gamma$--$Z$ interference,
while $W$-mediated contributions first appear at one-loop order.
Although QED alone does not generate a longitudinal parity-violating asymmetry,
QED radiative corrections modify $A_{\mathrm{LR}}$ through corrections to the
electromagnetic--weak interference and to the parity-conserving denominator.
A fully differential prediction additionally requires real photon radiation,
which is not considered in the present calculation. 
However, large logarithms due to soft or collinear real photon emission cancel in $A_\text{LR}$, and the remaining impact of real photon radiation depends on the experimental acceptance, which requires a simulation study beyond the scope of this work.

\begin{figure}
    \centering
    \includegraphics[width=1.00\linewidth]{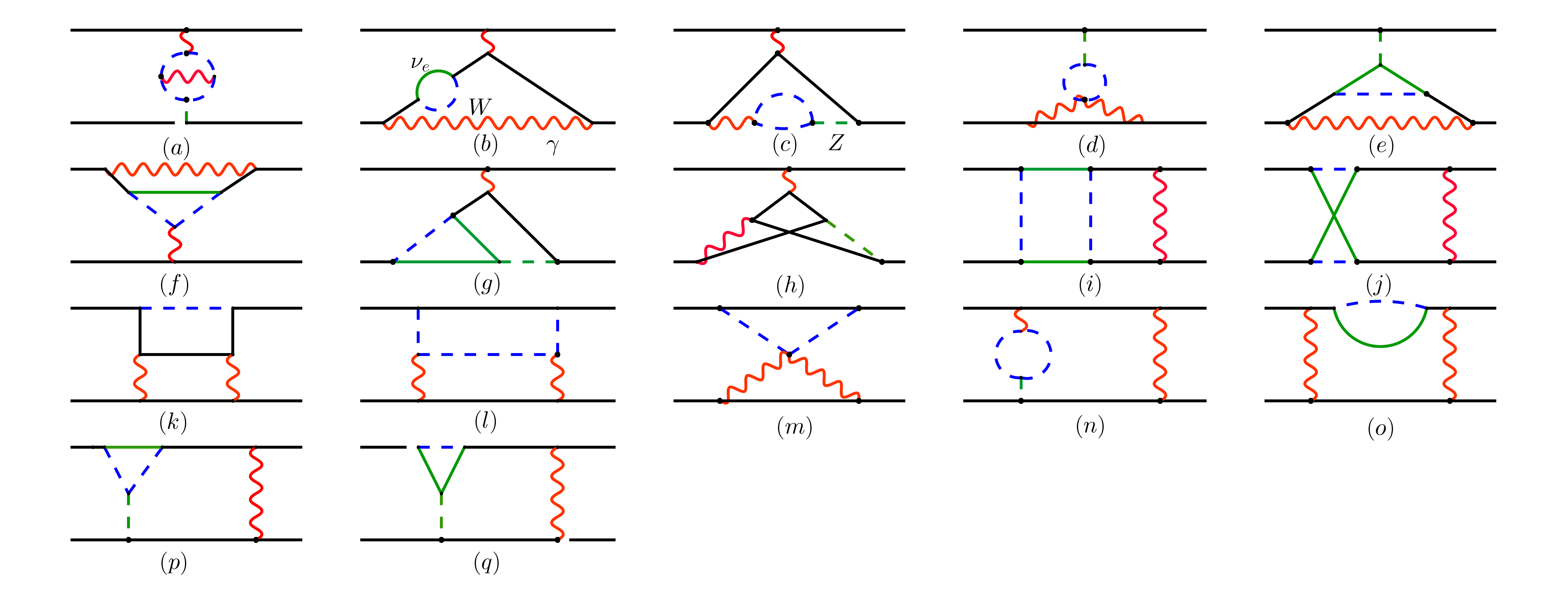}
    \caption{Examples of two-loop bosonic topologies, generated by \textsc{FeynGame}~\cite{Bundgen:2025utt}.
    Solid lines represent fermions, black labeling electrons and green labeling neutrinos.
    Red wavy lines represent photons.
    Dashed lines are EW gauge bosons, blue representing $W$-boson and green representing $Z$-boson.} 
    %Diagrams were drawn by FeynGame ~\cite{Bundgen:2025utt}.}
    \label{fig:2looptopos}
\end{figure}

The virtual NNLO contributions
considered here arise from the interference of EW bosonic two-loop amplitudes
with Born amplitudes and from products of bosonic one-loop amplitudes.
The two-loop bosonic amplitudes can be divided into self-energy,
vertex, and box diagrams, as well as one-particle-reducible contributions.
Representative genuine two-loop topologies are shown in Fig.~\ref{fig:2looptopos};
the reducible topologies are omitted from the figure but are included in our calculation.
%We exclude the purely fermionic loops since they have been studied in Ref.~\cite{Du:2019evk}.

\subsection{Regularization and Renormalization}
While the virtual corrections to $A_\text{LR}$ are infrared (IR) finite, soft and collinear divergencies must be regularized in individual loop diagrams. We accomplish this by introducing a small fictitious photon mass $m_\gamma$ and keeping the electron mass $m_e$ when necessary, respectively. 
The divergences lead to $\log m_{\gamma,e}^2$ terms that cancel in the final result, which has been verified explicitly.

Ultraviolet (UV) divergences are treated with dimensional regularization (dim-reg), where the spacetime dimension is generalized to
$D = 4 - 2 \epsilon$, and absorbed by appropriate counterterms.
Following the approach of Ref.~\cite{Du:2019evk},
we adopt the on-shell (OS) scheme for the renormalization of 
masses\footnote{Note that
the OS gauge-boson masses defined through the pole of the
radiatively corrected propagator differ from the mass definition
commonly used in experimental data listings, with the 
translation~\cite{Bardin:1988xt} given by
$$M_V = \frac{M_V^{\rm exp}}{\sqrt{1+
\left( \frac{\Gamma^{\rm exp}_V}{M_V^{\rm exp}} \right)^2}}
\hspace{30pt} (V=W,Z).$$}
and the electromagnetic coupling, and \MSbar\ renormalization for the weak mixing angle $\sin^2\theta_W$. 
Explicit expressions for the OS counterterms at NLO, which are needed for sub-loop 
renormalization and reducible contributions, can be found, {\em e.g.\/}, in 
Refs.~\cite{Denner:1991kt,Denner:2019vbn}.
The \MSbar\ renormalization of the weak mixing angle is motivated by the fact that the 
renormalization group running of $\hat{s}^2(\mu) \equiv \sin^2\hat\theta_W(\mu)$
absorbs the leading radiative corrections from fermion loops in the $t$-channel 
propagator~\cite{Marciano:1982mm}, as already mentioned in the introduction.
To ensure gauge invariance of the loop contributions,
the $W$-boson mass is derived from $M_Z$ and $\sin^2\hat\theta_W(0)$
by enforcing the relation $\sin^2\hat\theta_W(0) = 1-M_W^2/M_Z^2$ throughout, 
even in terms where we use the approximation $4\sin^2\theta_W \to 1$,
instead of being an independent input parameter.
In other words, the parameter $M_W$ in the loop integrals
does not correspond to the physical $W$-boson mass.

At NNLO, within the approximations used in this work (retaining only terms unsuppressed
by $1-4\sin^2\theta_W$ or proportional to $L^2$ with $L \equiv \ln M_Z^2/s$), only the 
counterterms for the $\gamma$--$Z$ mixing and the weak mixing angle are needed.
The construction of the former can be found in Ref.~\cite{Freitas:KIT}.
The NNLO \MSbar\ counterterm $\delta_{(2)}\hat{s}^2$ is determined by
matching its divergent part to the divergent part of the NNLO OS 
counterterm $\delta_{(2)}s_W$.
Here the subscript $(n)$ denotes the loop order.
In this construction, one needs to ensure that any appearances of $\sin^2\theta_W$ in 
the coefficients of the $\epsilon^{-2}$ and $\epsilon^{-1}$ poles are properly translated 
to $\hat{s}^2$. 
Furthermore, one has to account for the dependence of $\delta_{(1,2)}\hat{s}^2(\mu)$ on 
the renormalization scale $\mu$. 
In our calculation, we adopt a two-step process for this purpose.
First, the renormalization is carried out at $\mu=M_Z$,
where none of the SM fields are decoupled. 
Then we account for the change to $\mu=0$ by incorporating the appropriate contribution 
from renormalization-group (RG) running, expanded up to NNLO. 
Given that currently the electroweak contributions to the beta function and threshold 
corrections~\cite{Erler:2017knj} are included only up to ${\cal O}(\alpha)$, we performed 
the matching between the RG running and our new fixed-order calculation at this order. 
This means some terms that would otherwise be accounted for through $\hat s^2(0)$ are 
only included in our bosonic two-loop result without resummation.

\subsection{\texorpdfstring{$\gamma_5$}{gamma5} Prescription}
In dim-reg where $D=4-2\epsilon$,
the four-dimensional properties of $\gamma_5$ cannot all be maintained simultaneously.
Specifically, a fully anticommuting definition 
\begin{align}
\{\gamma^\mu,\gamma_5\}=0
\label{eq:antig5}
\end{align}
is incompatible with the cyclicity of the trace, 
or the non-vanishing four-dimensional identity
\begin{align}
\operatorname{Tr}\left(\gamma^\mu\gamma^\nu\gamma^\rho\gamma^\sigma\gamma_5\right)
=-4i\varepsilon^{\mu\nu\rho\sigma}.
\label{eq:cycltrg5}
\end{align}
As a result, calculations involving $\gamma_5$ in dim-reg require a rigorous and 
unambiguous prescription that properly handles the evanescent contributions, 
which occur when an ${\cal O}(\epsilon)$ mismatch in the algebra is multiplied 
by an UV-divergent integral.

In the present calculation, we adopt Na\"ive Dimensional Regularization (NDR)
supplemented by subtraction methods that allow us to perform $\gamma_5$-odd traces in 
four dimensions wherever necessary.
NDR assumes that the anticommutation relation~(\ref{eq:antig5}) holds and $\gamma_5^2=1$.
This is safe for two-loop self-energy diagrams, {\em e.g.\/} in 
Fig.~\ref{fig:2looptopos}$(a)$, since they do not receive any contributions from 
the Levi-Civita tensor due to the anti-symmetry of the latter.
For vertex diagrams, including ($b$-$h$) of Fig.~\ref{fig:2looptopos}, one can apply 
the vector/axial-vector projectors defined in Eqs.~(22-23) of Ref.~\cite{Freitas:2014hra},
which again ensures that Levi-Civita tensors do not contribute and NDR can be safely used.
The double-box topologies, ($i$-$m$) of Fig.~\ref{fig:2looptopos}, are free of 
UV-divergences.
Therefore, for these contributions the $\gamma_5$-sensitive traces 
can be evaluated directly in four dimensions. 
The identities from Ref.~\cite{Sirlin:1981pi} are useful for this purpose.

The more delicate cases are box diagrams containing UV-divergent sub-loops,
such as diagrams ($n$-$q$) of Fig.~\ref{fig:2looptopos}.
Each of these diagrams contains two $\gamma_5$-odd fermion chains in the polarized 
amplitudes, which result in non-vanishing contributions from Levi-Civita tensors.
Our prescription for them is to separate the UV divergence of the sub-loop from
the $\gamma_5$-sensitive fermion chains before performing the algebra.
Here, a fermion chain refers to the ordered product of Dirac matrices, spinors, and 
momenta associated with a given fermion line or the corresponding closed trace.

For self-energy insertions in boxes, {\em e.g.\/}, $(n)$ and $(o)$ of 
Fig.~\ref{fig:2looptopos},
the separation of UV poles from the fermion traces is rather straightforward.
One can project the sub-loop onto all possible orthogonal Lorentz structures of 
the self-energies in $D$ dimension before evaluating the traces, so that all information 
about the UV poles is captured in the coefficients of these Lorentz structures.
Alternatively, partial fraction decomposition at the amplitude level can separate
the sub-loop momentum $\ell_1$ from the fermion chains of the external electrons.
Both procedures reduce the contribution to a sum of dimensionally regulated
scalar sub-loop coefficients multiplying the outer fermion chains
that depend only on the external momenta $\{p_i\}$ and the outer-loop momentum $\ell_2$.
It therefore has precisely the same structure as the corresponding box diagram with a
self-energy counterterm insertion.
Hence, we can combine the two-loop diagram and the counterterm diagram
to construct a finite amplitude
\begin{equation}
\mathcal{M}_{\mathrm{ren}} =
\sum_i \int [d^D \ell_1 d^D \ell_2]\,
C^{\mathrm{sub}}_{i,\mathrm{ren}}(\ell_1,\ell_2,\{p\};D)
\left[ T_i^{\mathrm{out}}(\ell_2,\{p\};\gamma_5) \right]_{D=4},
\label{eq:gamma5-factorization}
\end{equation}
where $C^{\rm sub}_{i \rm ,ren} = C^{\rm sub}_{i \rm ,reg} + C^{\rm CT}_{i \rm ,reg}$ 
denotes the dimensionally regulated and renormalized coefficient of the sub-loop, 
while $T_i^{\rm out}$ denotes the remaining UV-finite outer loop of the box diagrams 
containing all fermion chains with $\gamma_5$.
Since the UV poles cancel in $C^{\rm sub}_{i \rm ,ren}$,
the Dirac algebra can be performed in four dimensions. 

Diagrams with UV-divergent vertex sub-loop, {\em e.g.\/}, ($p$) and ($q$) of 
Fig.~\ref{fig:2looptopos},
have more complicated Lorentz structures and introduce additional subtleties.
We treat these by generating local subtraction terms $\mathcal{M}_{\rm sub}$
that reproduce the UV behavior of the sub-loops of the genuine two-loop diagrams,
but are simpler to evaluate by themselves.
The difference $\mathcal{M}_{\text{UV-fin}} = \mathcal{M} - \mathcal{M}_{\rm sub}$ is 
UV-finite and can be calculated in four dimensions, whereas $\mathcal{M}_{\rm sub}$ can 
be integrated analytically and then added back and combined with the corresponding 
counterterm diagrams.
Pictorially, this process looks like 
\begin{equation*}
\hspace{-22pt}
{\hbox{\includegraphics[width=1.1\linewidth]{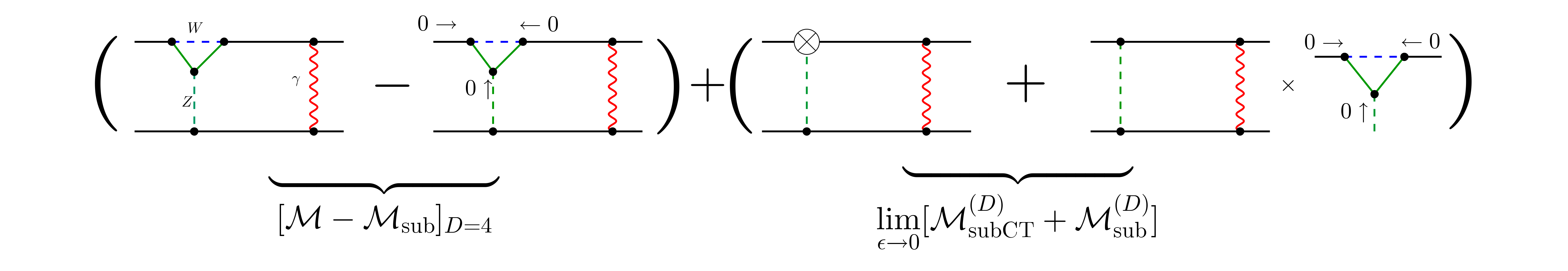}}}
\end{equation*}
The first diagram in each parenthesis is what needs to be computed for the physical 
process.
The second diagram in each parenthesis represents the additional term that is subtracted
and then added back to render each parenthesis UV-finite.
The arrows with zeros indicate that the subtracted amplitude is modified such that
the momentum flowing into each leg of the sub-loop vertex is set to zero without modifying 
anything else, {\em i.e.\/} the propagators of the sub-loop 
depend only on the inner loop momentum.
As a result, $\mathcal{M}_{\rm sub}$ constructed in this way can be factorized into two 
one-loop integrals, which can be evaluated analytically and naturally combined
with the box diagram with a sub-loop vertex counterterm.
Since each parenthesis is individually UV-finite, no dim-reg is needed and hence
no introduction of spurious finite terms that would otherwise arise from
evanescent $\mathcal{O}(D-4)$ contributions multiplying the UV poles.

\subsection{Analytic Evaluation of Integrals}
Throughout the calculation, we exploit the hierarchy of scales
\begin{align}
m_e^2 \ll Q^2 \ll M_V^2 \qquad (V = W,Z).
\label{eq:hierarchy}
\end{align}
As far as two-loop self-energies are concerned (see Fig.~\ref{fig:2looptopos}(a) for 
a sample diagram), the $\gamma$--$Z$ mixing self-energy is the only contribution enhanced 
in the sense discussed in the introduction. 
Its transverse part (the longitudinal part does not contribute) must be expanded to 
first order in $Q^2/M_V^2$, since it is multiplied by a photon propagator which is 
proportional to $Q^{-2}$.
This expansion can be carried out by performing a Taylor expansion at the integrand 
level~\cite{Freitas:KIT,Chen:2020xot}, and introduces derivatives of two-loop integrals, 
which can be reduced to analytically known master integrals using the relations in 
Refs.~\cite{Davydychev:1992mt,Weiglein:1993hd}.

The computations of vertex and box diagrams rely heavily on the Large Mass Expansion 
(LME) method~\cite{Smirnov:1990rz}.
Denoting the two loop momenta by $\ell_1$ and $\ell_2$, we assign to each momentum
either a hard scaling, $\ell_i\sim M_V$, or a soft scaling, $\ell_i\sim Q,m_e$.
For the diagrams considered in this paper, the soft-soft region where
$\{\ell_1,\ell_2\}\sim Q,m_e$ produces only scaleless integrals, which vanish in dim-reg,
or contributions with at least two local weak-interaction insertions
that scale like $G_F^2$ and are thus strongly suppressed.
This leaves us with three regions to consider:
\begin{center}
\begin{tabular}{c|cc}
Region & $\ell_1$ & $\ell_2$ \\ \hline
hard--hard & $M_V$ & $M_V$ \\
hard--soft & $M_V$ & $Q$   \\
soft--hard & $Q$   & $M_V$ \\
% soft--soft & $Q$   & $Q$
\end{tabular}
\end{center}
In each region, the integrand is expanded in a Taylor series, whose expansion parameters
are all quantities that are parametrically small under the corresponding scaling.
The expanded terms are then integrated over the full loop-momentum space,
after which the contributions from all regions are summed.
The sum of all relevant regions gives the asymptotic expansion of
the corresponding loop integral in $Q^2/M_V^2$ ($m_e^2/M_V^2$).
The results of this procedure are again two-loop vacuum integrals (from the hard-hard 
region since only the heavy masses are relevant) and products of one-loop integrals 
(from both the hard-hard and the hard-soft/soft-hard regions) whose analytic expressions 
are known~\cite{tHooft:1978jhc,Beenakker:1988jr,Ford:1992pn,Davydychev:1992mt}. 

After the LME, a second expansion for small $m_e/Q$ is performed. 
This is accomplished by simply setting $m_e \to 0$ in the IR-insensitive one-loop 
functions and in the coefficients of the one-loop functions.

\subsection{Computation Tools}
This calculation became possible through an extensive use of computer algebra tools.
The diagrams and amplitudes are generated with \textsc{FeynArts}~\cite{Hahn:2000kx}.
The Lorentz and Dirac algebra is evaluated independently using
\textsc{FeynCalc}~\cite{Shtabovenko:2016sxi,Shtabovenko:2023idz} and an in-house
\textsc{Mathematica} code.
The LME is implemented independently by different authors,
while \textsc{ASY2.1}~\cite{Jantzen:2012mw} is used as an additional check.
Integral reductions performed with our in-house codes are cross-checked using 
\textsc{FIRE7}~\cite{Smirnov:2025prc} and
\textsc{Kira~2.3}~\cite{Maierhofer:2017gsa,Klappert:2020nbg}.
All contributions to the final result have therefore been verified through
independent implementations by at least two authors.

\section{Results and Conclusions}
\label{sec:res}
The results of the two-loop purely bosonic contributions to $A_\text{LR}$ need to be combined 
with two-loop fermionic and mixed contributions~\cite{Du:2019evk}, as well as with 
the one-loop result~\cite{Czarnecki:1995fw}.
In order to elucidate the numerical significance of various contributions,
one has to specify the concrete perturbative scheme one is working with.
More precisely, one has to specify~\cite{Erler:2022ckm} what one considers the lowest 
order (tree level~\cite{Derman:1979zc}) expression.  
For example, identifying $\sin^2\theta_W$ in Eq.~\eqref{ALR} with the OS definition 
\eqref{sin2tonshell} or with the \MSbar\ weak mixing angle, $\sin^2\hat\theta(\mu)$, with 
$\mu = M_Z$ would lead to radiative corrections as large as 57\% or 39\%, respectively.
%This not only sheds doubt on the applicability of Feynman diagrammatic perturbation 
%theory, it also contrasts with the electroweak expansion parameter 
%$\alpha/(\pi\sin^2\theta_W) \sim 0.01$, from which one would expect smaller effects even 
%for contributions {\em not\/} suppressed by the $1-4\sin^2\theta_W$ factor discussed in 
%the introduction.
%Using instead the weak mixing angle in the \MSbar\ scheme, $\sin^2\hat\theta(\mu)$, with 
%the 't~Hooft scale chosen as $\mu = M_Z$, still results in corrections of 39\%.
The main reason for this can be traced to large logarithms $L \equiv \ln M_Z^2/s$ (or 
$\ln M_Z^2/m_f^2$ whenever $m_f^2 > s$).
Choosing instead $\mu = 0$ or $\mu = \sqrt{s} = 106$~MeV effectively re-sums such 
logarithms.
Thus, for our numerical analysis we write the lowest-order expression in the form
\begin{equation}
    A_\text{LR}^{LO} = \frac{G_F s}{\sqrt{8}\pi\alpha}\, \frac{y(1-y)}{(1-y+y^2)^2}\, 
    \underbrace{ \left[ 1-4\sin^2\hat\theta_W(0) \right]}_{0.04556} =
    4.08 \times 10^{-8},
\end{equation}
where the numerical value is for $y = 1/2$ where $A_\text{LR}$ is maximal.
In the following, we will present results for $y = 0.4$ together with the other input 
values used in Ref.~\cite{Erler:2022ckm}. 
Similar to what was done there, the electromagnetic couplings in the radiative 
corrections are expressed in terms of $\alpha(0)$, the electromagnetic coupling in the 
Thomson limit, and $\sin^2\hat\theta_W(0)$. 

Summarizing the one-loop result from Ref.~\cite{Czarnecki:1995fw}, 
the two-loop contributions from diagrams with at least one closed fermion loop from 
Ref.~\cite{Du:2019evk}, and the purely bosonic two-loop contributions obtained in the 
present work, we can write
\begin{align}
\label{effectiveQwe}
\nonumber
    Q_W^e &= (45.560 + 0.341_t + 0.774_B + 0.029_{(Q^2 \neq 0)} + 0.070_{FF} - 1.146_{FB} - 0.134_{BB})\times 10^{-3} \\
    &= 45.494 \times 10^{-3}.
\end{align}
The first three entries after the tree-level result represent the one-loop top-quark 
contribution including QCD corrections, the bosonic one-loop contribution, and the 
estimate for finite $Q^2$ effects~\cite{Erler:2022ckm}.
The next two terms are the two-loop contributions with two and one
closed fermion loop, respectively~\cite{Du:2019evk}.
Finally, the last term is the new result 
for the bosonic two-loop corrections presented in this paper, \emph{i.e.} the corrections 
without closed fermion loops that are enhanced by $(1-4\sin^2\theta_W)^{-1}$ or 
$\ln^2(M_Z^2/Q^2)$ relative to the tree-level contribution.

\begin{table}[t]
\centering
\begin{tabular}{|rcrr|}
\hline
loop order & amplitude class & result & uncertainty \\
\hline
tree & $Z$ & 45.560 & --- \\
1-loop & $(WW)$ & 5.689 & --- \\
1-loop & $(\gamma Z)$ & $-2.626$ & --- \\
1-loop & $(\gamma\gamma) Z$ & $-0.733$ & --- \\
2-loop & $(\gamma WW)$ & $-0.270$ & 0.023 \\
2-loop & $(\gamma\gamma)(WW)$ & $-0.038$ & 0.003 \\
2-loop & $(\gamma\gamma Z)$ & 0.174 & --- \\
2-loop & $(\gamma\gamma)(\gamma Z)$ & --- & 0.042 \\
2-loop & $(\gamma\gamma\gamma) Z$ & --- & 0.012 \\
3-loop & $(WWWW)$ & --- & $\ll 0.001$ \\
3-loop & $(\gamma\gamma WW)$ & --- & 0.019 \\
3-loop & $(\gamma\gamma\gamma Z)$ & --- & 0.009 \\
3-loop & $(\gamma\gamma)(\gamma WW)$ & --- & 0.005 \\
3-loop & $(\gamma\gamma)(\gamma\gamma Z)$ & --- & 0.002 \\
3-loop & $(\gamma\gamma\gamma)(WW)$ & --- & 0.001 \\
3-loop & $(\gamma\gamma\gamma)(\gamma Z)$ & --- & 0.001 \\
3-loop & $(\gamma\gamma\gamma\gamma)$ & --- & $\ll 0.001$ \\
4-loop & $(\gamma\gamma\gamma WW)$ & --- & 0.001 \\
\hline
all & & & 0.054 \\
\hline
\end{tabular}
\caption{Remaining uncertainties from purely bosonic contributions to $Q_W^e$ in units
of $10^{-3}$.
Each line corresponds to a term in the schematic expression \eqref{structure},
where we merged the pre-factors in brackets into one class, and to avoid clutter we also 
dropped here the powers of $\gamma$ in the denominators.
Notice that the three 1-loop results in the table do not add up to
$0.774_B$ in \eqref{effectiveQwe}.
The reason is that the numerical value for $(WW)$ only includes contributions from 
diagrams actually containing $W$ bosons (even a matching term already included in 
$\sin^2\hat\theta_W(0)$), but not Higgs or $ZZ$ box contributions because they enter
with the opposite sign so that dropping them here leads to a more
conservative uncertainty estimate.}
\label{errorsB}
\end{table}

In order to systematically estimate the uncertainty due to our approximation 
$\sin^2\theta_W = 1/4$ (in most of the terms), as well as from missing higher-order 
bosonic terms, we indicate the structure of the bosonic contributions up to three-loop 
order schematically in the form\footnote{For the uncertainty estimate we simply assume  
that all amplitudes are real.
Also note that the amplitude classes are defined such that each $\gamma$ may be replaced 
by a $Z$, but the resulting diagrams are sub-leading or even suppressed by additional 
powers of the Fermi constant.}
\begin{align} \nonumber
& \phantom+\ \frac{Z + (WW) + (\gamma Z)}{\gamma} - \frac{(\gamma\gamma) Z}{\gamma^2} \\
& + \frac{(\gamma WW) + (\gamma\gamma Z)}{\gamma} - \frac{(\gamma\gamma) [(WW) + 
(\gamma Z)]}{\gamma^2} - \left[ \frac{(\gamma\gamma\gamma)}{\gamma^2} - 
\frac{(\gamma\gamma)^2}{\gamma^3} \right] Z \nonumber \\
& + \frac{ (WWWW) + (\gamma\gamma WW) + (\gamma\gamma\gamma Z)}{\gamma} - 
\frac{(\gamma\gamma) [(\gamma WW) + (\gamma\gamma Z)]}{\gamma^2} \nonumber \\
& - \left[ \frac{(\gamma\gamma\gamma)}{\gamma^2} - \frac{(\gamma\gamma)^2}{\gamma^3} 
\right] [ (WW) + (\gamma Z)] - \left[ \frac{(\gamma\gamma\gamma\gamma)}{\gamma^2} - 2 
\frac{(\gamma\gamma)(\gamma\gamma\gamma)}{\gamma^3} + \frac{(\gamma\gamma)^3}{\gamma^4} 
\right] Z.
\label{structure}
\end{align}
Tree-level and one-loop terms appear in the first line, where expressions in parentheses 
represent classes of amplitudes.
For example, $(WW)$ not only refers to the $WW$ box diagram, but also to 
the $W$-loop contribution to $\gamma$-$Z$ mixing. 
These do not feature the $1-4\sin^2\theta_W$ suppression, in contrast to amplitudes 
containing even numbers of $Z$-boson-electron couplings,
such as the $\gamma Z$-box diagram $(\gamma Z)$.
It is also important to note that classes of diagrams involving both photons and massive 
bosons are enhanced by $L^n$, where $n$ is the number of photon lines connected fully
to the external electron lines in the vertex or box diagram. 
The two-photon amplitude $(\gamma\gamma)$ is not enhanced by a large logarithm but
by a numerically large coefficients $A \approx 6.9$, including a $\pi^2$-term related to 
the Sommerfeld enhancement. 
We conservatively assume that a factor of $A$ is contributed by each additional photon
in the amplitude class.
We can test this categorization by using it to anticipate the following numerical sizes 
for the amplitude classes associated with our NNLO bosonic result, where we identify the 
leading contribution $\propto L^2$ with the class $(\gamma\gamma Z)$, the sub-leading 
terms $\propto L$ with $(\gamma WW)$, and the remainder with the product class 
$(\gamma\gamma)(WW)$
\begin{align}
    \frac{(\gamma\gamma Z)}{\gamma} &\sim 
    - \frac{11}{6}\, L\, \frac{\alpha}{\pi} \frac{(\gamma Z)}{\gamma} \approx 
    0.151 \times 10^{-3} && [+0.174 \times 10^{-3}] \nonumber \\
    \frac{(\gamma WW)}{\gamma} &\sim 
    - \frac{11}{6}\, L\, \frac{\alpha}{\pi} \frac{(WW)}{\gamma} \approx 
    -0.328 \times 10^{-3} && [-0.270 \times 10^{-3}] \nonumber \\
    - \frac{(\gamma\gamma)(WW)}{\gamma^2} &\sim 
    - A\, \frac{\alpha}{\pi} \frac{(WW)}{\gamma} \approx 
    -0.092 \times 10^{-3} && [-0.038 \times 10^{-3}],
    \label{1lopestimates}
\end{align}
where we show [in brackets] the break-down of the result ($-0.134_{BB}\times 10^{-3}$) of
our 2-loop calculation 
(in the second and third line in the approximation $\sin^2\theta_W = 1/4$).
The pre-factor $11/6 = 3/2$ (from the $\gamma Z$ boxes) $+ 1/3$ (from the $Z$ contribution 
to the electron charge radius) multiplying $L$ is taken over from the one-loop result.

The uncertainties remaining from the bosonic corrections beyond our result, 
{\em i.e.\/} those neglected by using the $\sin^2\theta_W = 1/4$ and 
the leading-log approximations, are shown in Table~\ref{errorsB}.
The first two non-vanishing uncertainty entries account for the approximation
$\sin^2\theta_W = 1/4$, which we estimated by applying the scale 
factor $[4\sin^2\hat\theta_W(M_W)]^{-1} - 1$ to our respective results.
Indeed, comparing the non-logarithmic bosonic one-loop result
with the corresponding result obtained by using $\sin^2\theta_W = 1/4$ shows 
a 6\% difference, which is close to this scale factor.
All other uncertainty entries (including the other two 2-loop entries, which represent 
contributions from diagrams without $W$ bosons beyond the leading-log term in 
($\gamma\gamma Z$)) are computed using the same strategy as in \eqref{1lopestimates}.
As is also shown in the table, combining all error sources in quadrature (assuming no 
correlations or anticorrelations) we obtain $\pm 5.4\times 10^{-5}$.

In summary, our new result reduces the uncertainty from higher-order bosonic corrections  
from $\pm 3.8 \times 10^{-4}$ (without bosonic NNLO)\footnote{This is obtained by adding 
the estimates~\eqref{1lopestimates} and the other uncertainties in Table~\ref{errorsB} in 
quadrature, which is a safer and less arbitrary method than the variation of 
renormalization schemes or scales.} 
to $\pm 5.4\times 10^{-5}$ (after this work), or from about one third of 
the experimental precision goal for MOLLER~\cite{MOLLER:2014iki}, 
$\Delta Q_W^e \approx 0.0011$, to a negligible level. We provide an analytical expression of our bosonic two-loop result as an ancillary file {\tt Moller\_BB.res} in the arXiv submission. The file is readable in \textsc{Mathematica} and expressed in terms of the finite parts of the $B_0$ function \cite{tHooft:1978jhc} and the $I(1,1,1)$ function \cite{Davydychev:1992mt}. A detailed description of the symbols and notation can be found at the beginning of {\tt Moller\_BB.res}.

Discussions, refinements, and updates of some of the other contributions in 
Eq.~\eqref{effectiveQwe}, as well as a thorough discussion of the overall theoretical 
uncertainty in $Q_W^e$ are deferred to a subsequent publication.
A refinement is the complete low-energy effective field theory approach
to this computation, through which the large logarithms
$\ln M_{W,Z}^2/Q^2$ can be resummed \cite{Kollatzsch:2025pnp}.
Similarly, we defer a more careful evaluation of the impact of real-photon emission 
corrections to future work. 
Real photon radiation drops out in the fully inclusive asymmetry $A_{\rm LR}$, but needs 
to be taken into account when evaluating experimental acceptances and cuts 
(see \emph{e.g.} Ref.~\cite{Banerjee:2021qvi}).

\acknowledgments
This project has received funding from the European Research Council (ERC) under the 
European Union’s Horizon Europe research and innovation program under grant agreement 
No.~101142600 and under grant agreement No.~101292522 (project HADRON--2030), 
by the Deutsche Forschungsgemeinschaft (DFG, German Research Foundation) --- 
project number: 514321794 (CRC1660: Hadron and Nuclei as discovery tools), 
and by the U.S.\ National Science Foundation (Grant No.\ PHY-2412696).
LC and JE are also supported by the Cluster of Excellence 
{\em Precision Physics, Fundamental Interactions, and Structure of Matter\/} 
(PRISMA$^{++}$ EXC 2118/2) funded by the German Research Foundation (DFG) 
within the German Excellence Strategy (Project ID 390831469). 

\newpage

%%%%% references

%\renewcommand{\bibname}{References}
%\bibliographystyle{JHEP}
%\bibliography{./literature}

\begin{thebibliography}{10}

\bibitem{Erler:2014fqa}
J.~Erler, C.J.~Horowitz, S.~Mantry and P.A.~Souder, \emph{{Weak Polarized Electron Scattering}}, \href{https://doi.org/10.1146/annurev-nucl-102313-025520}{\emph{Ann. Rev. Nucl. Part. Sci.} {\bfseries 64} (2014) 269} [\href{https://arxiv.org/abs/1401.6199}{{\ttfamily 1401.6199}}].

\bibitem{Glashow:1961tr}
S.L.~Glashow, \emph{{Partial Symmetries of Weak Interactions}}, \href{https://doi.org/10.1016/0029-5582(61)90469-2}{\emph{Nucl. Phys.} {\bfseries 22} (1961) 579}.

\bibitem{Weinberg:1967tq}
S.~Weinberg, \emph{{A Model of Leptons}}, \href{https://doi.org/10.1103/PhysRevLett.19.1264}{\emph{Phys. Rev. Lett.} {\bfseries 19} (1967) 1264}.

\bibitem{Salam:1968rm}
A.~Salam, \emph{{Weak and Electromagnetic Interactions}}, \href{https://doi.org/10.1142/9789812795915_0034}{\emph{Conf. Proc. C} {\bfseries 680519} (1968) 367}.

\bibitem{Prescott:1979dh}
C.Y.~Prescott et~al., \emph{{Further Measurements of Parity Nonconservation in Inelastic electron Scattering}}, \href{https://doi.org/10.1016/0370-2693(79)91253-X}{\emph{Phys. Lett. B} {\bfseries 84} (1979) 524}.

\bibitem{Kumar:2013yoa}
K.S.~Kumar, S.~Mantry, W.J.~Marciano and P.A.~Souder, \emph{{Low Energy Measurements of the Weak Mixing Angle}}, \href{https://doi.org/10.1146/annurev-nucl-102212-170556}{\emph{Ann. Rev. Nucl. Part. Sci.} {\bfseries 63} (2013) 237} [\href{https://arxiv.org/abs/1302.6263}{{\ttfamily 1302.6263}}].

\bibitem{SLACE158:2005uay}
{\scshape SLAC E158} collaboration, \emph{{Precision measurement of the weak mixing angle in M\o ller scattering}}, \href{https://doi.org/10.1103/PhysRevLett.95.081601}{\emph{Phys. Rev. Lett.} {\bfseries 95} (2005) 081601} [\href{https://arxiv.org/abs/hep-ex/0504049}{{\ttfamily hep-ex/0504049}}].

\bibitem{Czarnecki:1998xc}
A.~Czarnecki and W.J.~Marciano, \emph{{Parity violating asymmetries at future lepton colliders}}, \href{https://doi.org/10.1142/S0217751X98001037}{\emph{Int. J. Mod. Phys. A} {\bfseries 13} (1998) 2235} [\href{https://arxiv.org/abs/hep-ph/9801394}{{\ttfamily hep-ph/9801394}}].

\bibitem{Erler:2004in}
J.~Erler and M.J.~Ramsey-Musolf, \emph{{The Weak mixing angle at low energies}}, \href{https://doi.org/10.1103/PhysRevD.72.073003}{\emph{Phys. Rev. D} {\bfseries 72} (2005) 073003} [\href{https://arxiv.org/abs/hep-ph/0409169}{{\ttfamily hep-ph/0409169}}].

\bibitem{MOLLER:2014iki}
{\scshape MOLLER} collaboration, \emph{{The MOLLER Experiment: An Ultra-Precise Measurement of the Weak Mixing Angle Using M\o ller Scattering}},  \href{https://arxiv.org/abs/1411.4088}{{\ttfamily 1411.4088}}.

\bibitem{Erler:2017knj}
J.~Erler and R.~Ferro-Hern{\'a}ndez, \emph{{Weak Mixing Angle in the Thomson Limit}}, \href{https://doi.org/10.1007/JHEP03(2018)196}{\emph{JHEP} {\bfseries 03} (2018) 196} [\href{https://arxiv.org/abs/1712.09146}{{\ttfamily 1712.09146}}].

\bibitem{Erler:2024lds}
J.~Erler, R.~Ferro-Hernandez and S.~Kuberski, \emph{{Theory-Driven Evolution of the Weak Mixing Angle}}, \href{https://doi.org/10.1103/PhysRevLett.133.171801}{\emph{Phys. Rev. Lett.} {\bfseries 133} (2024) 171801} [\href{https://arxiv.org/abs/2406.16691}{{\ttfamily 2406.16691}}].

\bibitem{Davoudiasl:2012ag}
H.~Davoudiasl, H.-S.~Lee and W.J.~Marciano, \emph{{`Dark' Z implications for Parity Violation, Rare Meson Decays, and Higgs Physics}}, \href{https://doi.org/10.1103/PhysRevD.85.115019}{\emph{Phys. Rev. D} {\bfseries 85} (2012) 115019} [\href{https://arxiv.org/abs/1203.2947}{{\ttfamily 1203.2947}}].

\bibitem{Pati:1974yy}
J.C.~Pati and A.~Salam, \emph{{Lepton Number as the Fourth Color}}, \href{https://doi.org/10.1103/PhysRevD.10.275}{\emph{Phys. Rev. D} {\bfseries 10} (1974) 275}.

\bibitem{Czarnecki:1995fw}
A.~Czarnecki and W.J.~Marciano, \emph{{Electroweak radiative corrections to polarized Moller scattering asymmetries}}, \href{https://doi.org/10.1103/PhysRevD.53.1066}{\emph{Phys. Rev. D} {\bfseries 53} (1996) 1066} [\href{https://arxiv.org/abs/hep-ph/9507420}{{\ttfamily hep-ph/9507420}}].

\bibitem{Du:2019evk}
Y.~Du, A.~Freitas, H.H.~Patel and M.J.~Ramsey-Musolf, \emph{{Parity-Violating M{\o}ller Scattering at Next-to-Next-to-Leading Order: Closed Fermion Loops}}, \href{https://doi.org/10.1103/PhysRevLett.126.131801}{\emph{Phys. Rev. Lett.} {\bfseries 126} (2021) 131801} [\href{https://arxiv.org/abs/1912.08220}{{\ttfamily 1912.08220}}].

\bibitem{Bundgen:2025utt}
L.~B{\"u}ndgen, R.V.~Harlander, S.Y.~Klein and M.C.~Schaaf, \emph{{FeynGame 3.0}}, \href{https://doi.org/10.1016/j.cpc.2025.109662}{\emph{Comput. Phys. Commun.} {\bfseries 314} (2025) 109662} [\href{https://arxiv.org/abs/2501.04651}{{\ttfamily 2501.04651}}].

\bibitem{Bardin:1988xt}
D.Y.~Bardin, A.~Leike, T.~Riemann and M.~Sachwitz, \emph{{Energy Dependent Width Effects in e+ e- Annihilation Near the Z Boson Pole}}, \href{https://doi.org/10.1016/0370-2693(88)91627-9}{\emph{Phys. Lett. B} {\bfseries 206} (1988) 539}.

\bibitem{Denner:1991kt}
A.~Denner, \emph{{Techniques for calculation of electroweak radiative corrections at the one loop level and results for W physics at LEP-200}}, \href{https://doi.org/10.1002/prop.2190410402}{\emph{Fortsch. Phys.} {\bfseries 41} (1993) 307} [\href{https://arxiv.org/abs/0709.1075}{{\ttfamily 0709.1075}}].

\bibitem{Denner:2019vbn}
A.~Denner and S.~Dittmaier, \emph{{Electroweak Radiative Corrections for Collider Physics}}, \href{https://doi.org/10.1016/j.physrep.2020.04.001}{\emph{Phys. Rept.} {\bfseries 864} (2020) 1} [\href{https://arxiv.org/abs/1912.06823}{{\ttfamily 1912.06823}}].

\bibitem{Marciano:1982mm}
W.J.~Marciano and A.~Sirlin, \emph{{Radiative Corrections to Atomic Parity Violation}}, \href{https://doi.org/10.1103/PhysRevD.27.552}{\emph{Phys. Rev. D} {\bfseries 27} (1983) 552}.

\bibitem{Freitas:KIT}
A.~Freitas, \emph{{Fermionische Zwei-Schleifen-Beitr\"{a}ge zum Myon-Zerfall}},  Universit\"{a}t Karlsruhe, 1999, \href{https://www.itp.kit.edu/prep/diploma/PSFiles/diploma-2-1999.ps.gz}{https://www.itp.kit.edu/prep/diploma/PSFiles/diploma-2-1999.ps.gz}.

\bibitem{Freitas:2014hra}
A.~Freitas, \emph{{Higher-order electroweak corrections to the partial widths and branching ratios of the Z boson}}, \href{https://doi.org/10.1007/JHEP04(2014)070}{\emph{JHEP} {\bfseries 04} (2014) 070} [\href{https://arxiv.org/abs/1401.2447}{{\ttfamily 1401.2447}}].

\bibitem{Sirlin:1981pi}
A.~Sirlin, \emph{{A Class of Useful Identities Involving Correlated Direct Products of $\gamma$ Matrices}}, \href{https://doi.org/10.1016/0550-3213(81)90195-4}{\emph{Nucl. Phys. B} {\bfseries 192} (1981) 93}.

\bibitem{Chen:2020xot}
L.~Chen and A.~Freitas, \emph{{Mixed EW-QCD leading fermionic three-loop corrections at $\mathcal{O}(\alpha_s\alpha^2)$ to electroweak precision observables}}, \href{https://doi.org/10.1007/JHEP03(2021)215}{\emph{JHEP} {\bfseries 03} (2021) 215} [\href{https://arxiv.org/abs/2012.08605}{{\ttfamily 2012.08605}}].

\bibitem{Davydychev:1992mt}
A.I.~Davydychev and J.B.~Tausk, \emph{{Two-loop self-energy diagrams with different masses and the momentum expansion}}, \href{https://doi.org/10.1016/0550-3213(93)90338-P}{\emph{Nucl. Phys. B} {\bfseries 397} (1993) 123}.

\bibitem{Weiglein:1993hd}
G.~Weiglein, R.~Scharf and M.~Bohm, \emph{{Reduction of general two-loop self-energies to standard scalar integrals}}, \href{https://doi.org/10.1016/0550-3213(94)90325-5}{\emph{Nucl. Phys. B} {\bfseries 416} (1994) 606} [\href{https://arxiv.org/abs/hep-ph/9310358}{{\ttfamily hep-ph/9310358}}].

\bibitem{Smirnov:1990rz}
V.A.~Smirnov, \emph{{Asymptotic expansions in limits of large momenta and masses}}, \href{https://doi.org/10.1007/BF02102092}{\emph{Commun. Math. Phys.} {\bfseries 134} (1990) 109}.

\bibitem{tHooft:1978jhc}
G.~'t~Hooft and M.J.G.~Veltman, \emph{{Scalar One Loop Integrals}}, \href{https://doi.org/10.1016/0550-3213(79)90605-9}{\emph{Nucl. Phys. B} {\bfseries 153} (1979) 365}.

\bibitem{Beenakker:1988jr}
W.~Beenakker and A.~Denner, \emph{{Infrared Divergent Scalar Box Integrals With Applications in the Electroweak Standard Model}}, \href{https://doi.org/10.1016/0550-3213(90)90636-R}{\emph{Nucl. Phys. B} {\bfseries 338} (1990) 349}.

\bibitem{Ford:1992pn}
C.~Ford, I.~Jack and D.R.T.~Jones, \emph{{The Standard model effective potential at two loops}}, \href{https://doi.org/10.1016/0550-3213(92)90165-8}{\emph{Nucl. Phys. B} {\bfseries 387} (1992) 373} [\href{https://arxiv.org/abs/hep-ph/0111190}{{\ttfamily hep-ph/0111190}}].

\bibitem{Hahn:2000kx}
T.~Hahn, \emph{{Generating Feynman diagrams and amplitudes with FeynArts 3}}, \href{https://doi.org/10.1016/S0010-4655(01)00290-9}{\emph{Comput. Phys. Commun.} {\bfseries 140} (2001) 418} [\href{https://arxiv.org/abs/hep-ph/0012260}{{\ttfamily hep-ph/0012260}}].

\bibitem{Shtabovenko:2016sxi}
V.~Shtabovenko, R.~Mertig and F.~Orellana, \emph{{New Developments in FeynCalc 9.0}}, \href{https://doi.org/10.1016/j.cpc.2016.06.008}{\emph{Comput. Phys. Commun.} {\bfseries 207} (2016) 432} [\href{https://arxiv.org/abs/1601.01167}{{\ttfamily 1601.01167}}].

\bibitem{Shtabovenko:2023idz}
V.~Shtabovenko, R.~Mertig and F.~Orellana, \emph{{FeynCalc 10: Do multiloop integrals dream of computer codes?}}, \href{https://doi.org/10.1016/j.cpc.2024.109357}{\emph{Comput. Phys. Commun.} {\bfseries 306} (2025) 109357} [\href{https://arxiv.org/abs/2312.14089}{{\ttfamily 2312.14089}}].

\bibitem{Jantzen:2012mw}
B.~Jantzen, A.V.~Smirnov and V.A.~Smirnov, \emph{{Expansion by regions: revealing potential and Glauber regions automatically}}, \href{https://doi.org/10.1140/epjc/s10052-012-2139-2}{\emph{Eur. Phys. J. C} {\bfseries 72} (2012) 2139} [\href{https://arxiv.org/abs/1206.0546}{{\ttfamily 1206.0546}}].

\bibitem{Smirnov:2025prc}
A.V.~Smirnov and M.~Zeng, \emph{{FIRE 7: Automatic Reduction with Modular Approach}},  \href{https://arxiv.org/abs/2510.07150}{{\ttfamily 2510.07150}}.

\bibitem{Maierhofer:2017gsa}
P.~Maierh{\"o}fer, J.~Usovitsch and P.~Uwer, \emph{{Kira{\textemdash}A Feynman integral reduction program}}, \href{https://doi.org/10.1016/j.cpc.2018.04.012}{\emph{Comput. Phys. Commun.} {\bfseries 230} (2018) 99} [\href{https://arxiv.org/abs/1705.05610}{{\ttfamily 1705.05610}}].

\bibitem{Klappert:2020nbg}
J.~Klappert, F.~Lange, P.~Maierh{\"o}fer and J.~Usovitsch, \emph{{Integral reduction with Kira 2.0 and finite field methods}}, \href{https://doi.org/10.1016/j.cpc.2021.108024}{\emph{Comput. Phys. Commun.} {\bfseries 266} (2021) 108024} [\href{https://arxiv.org/abs/2008.06494}{{\ttfamily 2008.06494}}].

\bibitem{Erler:2022ckm}
J.~Erler, R.~Ferro-Hern{\'a}ndez and A.~Freitas, \emph{{Hadronic effects in M{\o}ller scattering at NNLO}}, \href{https://doi.org/10.1007/JHEP08(2022)183}{\emph{JHEP} {\bfseries 08} (2022) 183} [\href{https://arxiv.org/abs/2202.11976}{{\ttfamily 2202.11976}}].

\bibitem{Derman:1979zc}
E.~Derman and W.J.~Marciano, \emph{{Parity Violating Asymmetries in Polarized Electron Scattering}}, \href{https://doi.org/10.1016/0003-4916(79)90095-2}{\emph{Annals Phys.} {\bfseries 121} (1979) 147}.

\bibitem{Kollatzsch:2025pnp}
S.~Kollatzsch, D.~Moreno, D.~Radic and A.~Signer, \emph{{Parity violation in M{\o}ller scattering within low-energy effective field theory}}, \href{https://doi.org/10.1007/JHEP09(2025)196}{\emph{JHEP} {\bfseries 09} (2025) 196} [\href{https://arxiv.org/abs/2507.17652}{{\ttfamily 2507.17652}}].

\bibitem{Banerjee:2021qvi}
P.~Banerjee, T.~Engel, N.~Schalch, A.~Signer and Y.~Ulrich, \emph{{M{\o}ller scattering at NNLO}}, \href{https://doi.org/10.1103/PhysRevD.105.L031904}{\emph{Phys. Rev. D} {\bfseries 105} (2022) L031904} [\href{https://arxiv.org/abs/2107.12311}{{\ttfamily 2107.12311}}].

\end{thebibliography}

\providecommand{\href}[2]{#2}\begingroup\raggedright\endgroup

\end{document}